\documentclass[aps,pra,amssymb,superscriptaddress,10pt,tightenlines,twocolumn,showpacs]{revtex4-1}

\usepackage[english]{babel}
\usepackage{graphicx}
\usepackage{dcolumn}
\usepackage{bm}
\usepackage{color}
\usepackage{subfigure}
\usepackage[ansinew]{inputenc}
\usepackage{amsfonts}
\usepackage{amsmath}

\def\bra#1{\langle #1 |}
\def\ket#1{| #1 \rangle}

\def\e{\mathrm{e}}

\def\e{\mathrm{e}}
\def\i{\mathrm{i}}

\renewcommand{\Re}{\mathop{\text{Re}}\nolimits}

\begin{document}

\title{Bound states and entanglement generation in waveguide quantum electrodynamics}

\author{Paolo Facchi}
\affiliation{Dipartimento di Fisica and MECENAS, Universit\`{a} di Bari, I-70126 Bari, Italy}
\affiliation{INFN, Sezione di Bari, I-70126 Bari, Italy}

\author{M. S. Kim}
\affiliation{QOLS, Blackett Laboratory, Imperial College London, London SW7 2AZ, United Kingdom.}

\author{Saverio Pascazio}
\affiliation{Dipartimento di Fisica and MECENAS, Universit\`{a} di Bari, I-70126 Bari, Italy}
\affiliation{INFN, Sezione di Bari, I-70126 Bari, Italy}

\author{Francesco V. Pepe}
\affiliation{Museo Storico della Fisica e Centro Studi e Ricerche ``Enrico Fermi'', I-00184 Roma, Italy}
\affiliation{INFN, Sezione di Bari, I-70126 Bari, Italy}

\author{Domenico Pomarico}
\affiliation{Dipartimento di Fisica and MECENAS, Universit\`{a} di Bari, I-70126 Bari, Italy}

\author{Tommaso Tufarelli}
\affiliation{School of Mathematical Sciences, University of Nottingham,Nottingham NG7 2RD, United Kingdom}

\date{\today}

\begin{abstract}
{\noindent}We investigate the behavior of two quantum emitters (two-level atoms) embedded in a linear waveguide, in a quasi-one-dimensional configuration. Since the atoms can emit, absorb and reflect radiation, the pair can spontaneously relax towards an entangled bound state, under conditions in which a single atom would instead decay. Exploting the resolvent formalism, we analyze the properties of these bound states, which occur for resonant values of the interatomic distance, and discuss their relevance with respect to entanglement generation. The stability of such states close to the resonance is studied, as well as the properties of non-resonant bound states, whose energy is below the threshold for photon propagation.
\end{abstract}


\maketitle

\section{Introduction} 
An excited atom in free space unavoidably decays towards its ground state through  spontaneous emission. Boundary conditions and artificial dimensional reduction can drastically modify the picture, providing situations in which the decay is enhanced, inhibited or even completely hindered~\cite{QED1,QED2,QED3,decay1,decay2,decay3,decay4,decay5,decay6}. While confinement in optical cavities has long been a common way to study the effects of geometry~\cite{cavity1,cavity2,cavity3}, one-dimensional systems have recently emerged as another promising stage for the observation of interesting quantum electrodynamics (QED) phenomena. Nowadays a variety of quantum emitters (atoms for brevity) can be coupled to quasi one-dimensional fields such as waveguides, optical fibers and microwave transmission lines~\cite{onedim1,onedim2,onedim3,onedim4,onedim5,onedim6,onedim7}. Alternatively, the effective reduction to one dimension can be obtained by tightly focusing photons~\cite{focused1,focused2,focused3}. These impressive experimental advances have opened the way to unexplored nonperturbative regimes of QED, and have motivated work on the interaction between atoms and waveguides in different geometries \cite{cirac1,cirac2,kimble1,kimble2,ck,refereeA3,refereeA4,refereeB1}.

In this context, an interesting problem is the study of atoms in semi-infinite linear waveguides, where one end of the guide behaves as a perfect mirror~\cite{semiinfinite1,semiinfinite2,leo5}. For selected values of the atom-mirror distance a nontrivial bound state exists, in which the  probability of atomic excitation is finite, even when photons emitted through spontaneous decay can propagate in the guide~\cite{mirror1,mirror2}. The optical path between the atom and the mirror is crucial for the existence of this kind of resonance. It is worth noting that even a single atom exhibits a mirror-like behavior in one dimension~\cite{focused2,focused3,atomrefl1,atomrefl2,atomrefl3,NJP}. One may thus consider the interaction of two atoms, mediated by the exchange of photons propagating in one dimension, and exploit the dual behavior of each atom as both an emitter and a mirror. Such interaction can give rise to stable configurations in which the atoms display significant entanglement, while the field is confined between the atoms and does not propagate \cite{NJP,refereeA1,refereeA2}. Besides the fundamental interest of few-body QED in quasi 1D geometries, where non-Markovian effects easily come into play \cite{TKC}, such a system is thus interesting from the point of view of generating entanglement, an important resource in Quantum Information, by {\it relaxation}. 
Indeed, if a bound state exists in which the two atoms are entangled, an initially factorized atomic state can spontaneously relax towards a state with finite entanglement. Relaxation occurs after an initial transient in which photon exchange builds up quantum correlations. Differently from other methods of entanglement generation in waveguide-QED \cite{barangio2}, this process would not require a continuous pumping of energy into the system, and would ideally provide a constant entanglement in time after the initial transient.

In this paper we show how the properties of bound and quasi-bound entangled states in waveguide-QED can be studied in great depth and generality by exploiting the resolvent formalism \cite{CT,NNP}. Studying the resolvent, one notices the presence of a number of poles in the so-called \textit{complex-energy plane}. Each pole can be associated with a (generally unstable) state, and the imaginary part of a pole is proportional to the inverse lifetime of the state. This allows us to immediately identify a favorable situation for entanglement generation by relaxation: we need one of these poles to be a long-lived entangled state (i.e., the pole must have a negligible imaginary part), while the remaining poles must be fast-decaying states. Under such conditions, a separable atomic state would quickly relax onto an entangled metastable state. While the metastable state will eventually decay due to losses and imperfections, our analysis allows to clearly identify the relevant timescales of the problem. Thus, we can give a clear indication of what degree of losses and imperfections a given system is able to tolerate while still allowing the generation of long-lived entanglement. Importantly, our formalism automatically takes into account a number of physical effect that are often neglected: these include non-markovian effects, time-delay (due to the finite propagation speed of photons), threshold effects (due to the presence of either high- or low- frequency cutoffs in the dispersion relation). Even though, for definiteness, we focus for the most part on the dispersion relation typical of rectangular waveguides, we will also outline how a wide class of physically relevant dispersion relations can be tackled within the same framework.

Our paper is organized as follows. In section \ref{model} we introduce the Hamiltonian of our model, and illustrate how a suitable choice of the inter-atomic distance gives rise to entangled bound states above the frequency threshold for photon propagation. Section~\ref{resolvent} is devoted to the study of poles in the complex-energy plane, which allows us to extract crucial information relevant to the entanglement-by-relaxation protocol. In section~\ref{offres} we extend our analysis to off-resonant bound states, whose energy is below the low-frequency cutoff of the waveguide. We outline in section~\ref{sec:general} how our study can be generalized in a straightforward manner to any dispersion relation that satisfies appropriate conditions. Finally, we draw our conclusions in section~\ref{conclusions}.
\begin{figure}[t!]
	\includegraphics[width=.85\linewidth]{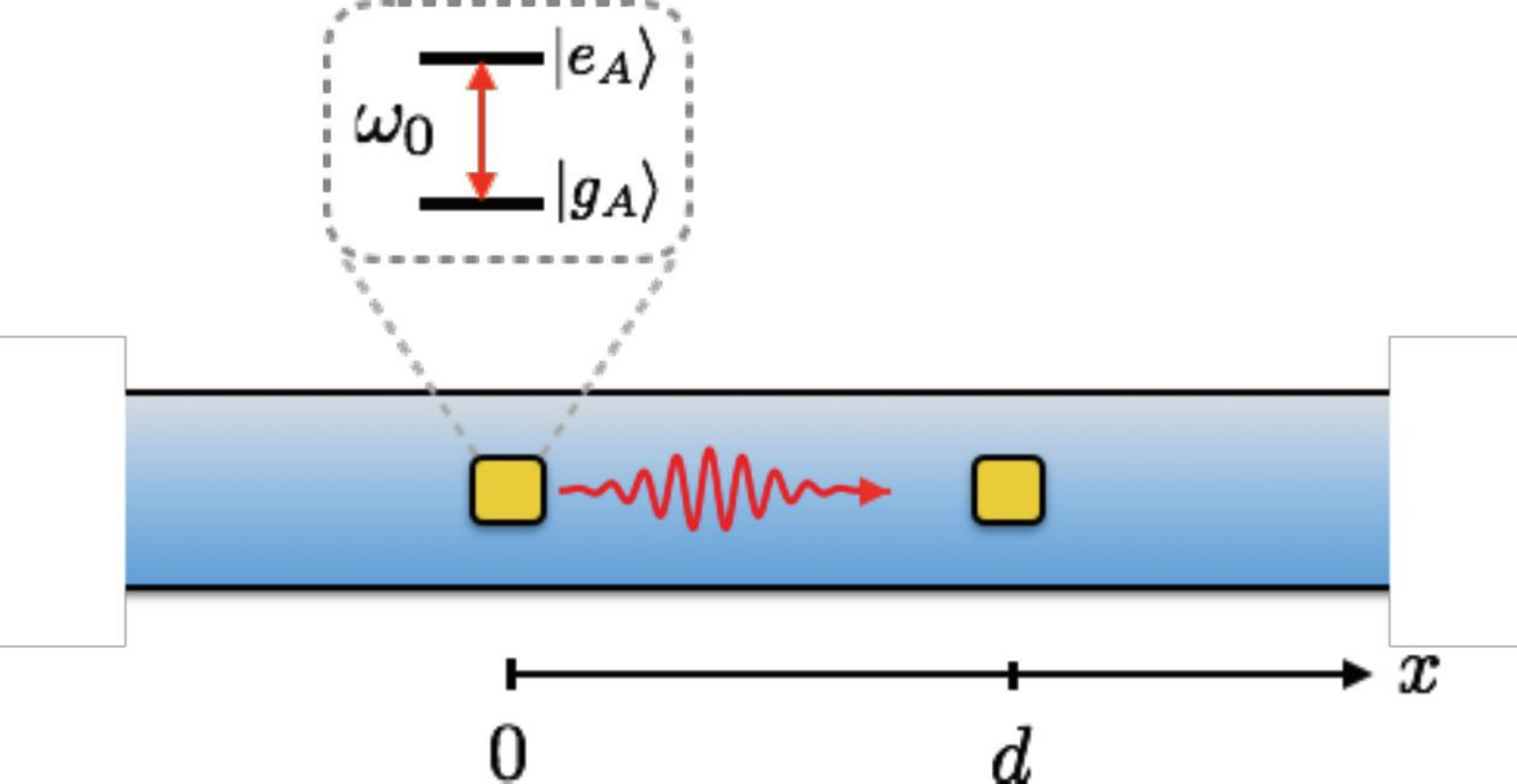}
	\caption{Two two-level atoms are placed at relative distance $d$ in a one-dimensional waveguide, with propagation direction along the $x$ axis. Both atoms possess the same internal structure (for brevity we only sketch the levels of emitter $A$) and interact through the mediation of waveguide photons. 
	The waveguide is characterized by its one-dimensional photon dispersion relation $\omega(k)$, with $k$ being the photon momentum. We first focus on the $\mathrm{TE}_{1,0}$ mode of an infinite waveguide of rectangular cross section, and then generalize to a wide class of one-dimensional dispersion relations in Sec.\ \ref{sec:general}.
\label{Fig1}}
\end{figure}

\section{The model}\label{model}
We describe the dynamics of two two-level atoms $A$ and $B$, situated in an infinite waveguide of rectangular cross section, with sides $L_y<L_z$, see Fig.~\ref{Fig1}. When longitudinal propagation occurs with long wavelength compared to the transverse size, interaction between atoms and field can be reduced to a coupling with the lowest-cutoff-energy $\mathrm{TE}_{1,0}$ mode, in which the electric field vibrates along the $z$ direction and has a sine modulation in the $y$ direction~\cite{jackson}. In this situation, the electromagnetic field is effectively scalar and massive. 
The interacting atoms and photons are described, in dipolar and rotating wave approximations, by the Hamiltonian
\begin{eqnarray}
H&=& H_0+\lambda V\nonumber\\
& = & \omega_0 (\ket{e_A}\bra{e_A} + \ket{e_B}\bra{e_B}) + \int d k\, \omega(k) b^{\dagger}(k) b(k) \nonumber \\ 
& & + \lambda \int \!\frac{d k}{\omega(k)^{1/2}} \Bigl[ \ket{e_A}\bra{g_A} b(k) + \ket{g_A}\bra{e_A} b^{\dagger}(k) \nonumber \\
& & \qquad + \ket{e_B}\bra{g_B} b(k) \e^{\i kd} + \ket{g_B}\bra{e_B} b^{\dagger}(k) \e^{-\i kd} \Bigr], \quad
\label{hamiltonian}
\end{eqnarray} 
where $\omega_0$ is the bare energy separation between the atomic ground $\ket{g}$ and first-excited states $\ket{e}$, $\lambda$ is the coupling constant (see Appendix B), $d$ is the $A$-$B$ distance, $\omega(k)$ is the photon dispersion relation, and $b(k)$ ($b^{\dagger}(k)$) is the annihilation (creation) field operator, satisfying the canonical commutation relation $[b(k),b^{\dagger}(k')]=\delta(k-k')$. Henceforth, we will focus on the dispersion $\omega(k)=\sqrt{k^2+M^2}$ of the $\mathrm{TE}_{1,0}$ mode in the waveguide, characterized by a mass $M\propto L_y^{-1}$. However, as discussed in section \ref{sec:general}, our approach is applicable to a wide class of one-dimensional dispersion relations.
The effective mass $M$ provides a natural cutoff to the coupling. The Hamiltonian~(\ref{hamiltonian}) commutes with the excitation number 
\begin{equation}\label{number}
\mathcal{N} = \mathcal{N}_{\mathrm{at}} 
+ \int d k\, b^{\dagger}(k) b(k),
\end{equation}
where $\mathcal{N}_{\mathrm{at}}= \ket{e_A}\bra{e_A} + \ket{e_B}\bra{e_B}$ is the atomic excitation number. 
The $\mathcal{N}=0$ sector is 1-dimensional and is spanned by the bare ground state $\ket{g_A,g_B;\mathrm{vac}}$. 
We shall focus instead on the dynamics in the $\mathcal{N}=1$ sector, where the states read
\begin{equation}\label{wavefunction}
\ket{\psi} =  \big(c_A \ket{e_A,g_B}+ c_B \ket{g_A,e_B}\big) \otimes \ket{\mathrm{vac}} + \ket{g_A,g_B} \otimes \ket{\varphi}
\end{equation}
where $\ket{\varphi}:=\int d k\, \varphi(k) b^{\dagger}(k)\ket{\mathrm{vac}}$ is a one-photon state,
and $|c_A|^2+|c_B|^2+\int d k |\varphi(k)|^2=1$.

In the small-coupling regime, an isolated excited atom with $\omega_0\gtrsim M$ would decay to the ground state. We shall demonstrate that, when two atoms are considered, a resonance effect emerges, yielding a bound state. Using the expansion~(\ref{wavefunction}) the eigenvalue equation, $H\ket{\psi}=E\ket{\psi}$, reads 
\begin{eqnarray}
E c_A & = & \omega_0 c_A + \lambda \int d k \frac{\varphi(k)}{\omega(k)^{1/2}}, \label{eigenvalueA} \\ 
E c_B & = & \omega_0 c_B + \lambda \int d k \frac{\varphi(k) \e^{\i kd}}{\omega(k)^{1/2}}, \label{eigenvalueB} \\ 
\varphi(k) & = &  \frac{\lambda}{\omega(k)^{1/2}} \frac{c_A + c_B \e^{-\i kd}}{E - \omega(k)}. \label{eigenvalueF}
\end{eqnarray}
The field amplitude $\varphi(k)$ has two simple poles at $k=\pm \bar{k} = \pm\sqrt{E^2-M^2}$. Thus, when $E>M$, the integrals in~(\ref{eigenvalueA})-(\ref{eigenvalueB}) are finite only if $c_A+c_B \e^{\pm i \bar{k} d}=0$, yielding $\bar{k}  d= n\pi$ for positive integers $n$.
This implies that a bound state can exist only for discrete values of the interatomic distance $d$. Moreover, in the first component of such an eigenstate (\ref{wavefunction}), the atoms are in a maximally entangled (singlet or triplet) state, namely $c_A=(-1)^{n+1}c_B$. To determine the distances at which the bound state exists, let us first compute the energy eigenvalue, which after the resonance condition is the solution of
\begin{eqnarray}\label{eigenvalue}
E & = & \omega_0 + \lambda^2 \int d k \frac{1-(-1)^n \e^{-\i kd}}{\omega(k)(E-\omega(k))} \nonumber \\
& = & \omega_0 + \frac{2 \lambda^2}{M} \left[ 1 + O\!\left( \frac{E-M}{M} \right) + O\!\left( \frac{\e^{-Md}}{\sqrt{Md}} \right) \right]. 
\end{eqnarray}
Corrections in the second line are negligible if $|\omega_0-M|\ll M$. This will result as a special case of the ensuing analysis of the complex poles of the resolvent. [See Eq.~(\ref{poles}) and following ones.]
 Thus for large $M$, 
a bound state with $E>M$ is present only if the distance $d$ takes one of the discrete and equally spaced values
\begin{equation}\label{kbar}
d_n=\frac{n\pi}{\bar{k}}, \quad \text{with} \quad \bar{k}:= \sqrt{\left( \omega_0+\frac{2 \lambda^2}{M} \right)^2 -M^2},
\end{equation}
and if the wavenumber $\bar{k}$ is real ($\omega_0>M-2\lambda^2/M$). The properties of states with $E<M$, to which an imaginary wavenumber can be associated, will be discussed in Section \ref{offres}.

To complete the characterization of the bound state, we shall analyze the atomic populations and the field energy density. The former can be immediately obtained using the normalization condition on the states~(\ref{wavefunction}) as
\begin{equation}
1=2|c_A^{(n)}|^2 \left(1 + \lambda^2 \int d k \frac{1-(-1)^n \cos(kd_n)}{\omega(k)(E-\omega(k))^2} \right),
\end{equation}
Where we use the shorthands $c_A^{(n)},c_B^{(n)}$ to indicate the coefficients of the bound state with $d=d_n$. Retaining only the highest order in $M$ and defining $p_n:=|c_A^{(n)}|^2+|c_B^{(n)}|^2$ as the probability associated to the $\mathcal{N}_{\mathrm{at}}=1$ sector, one gets
\begin{equation}\label{probability}
p_n \simeq \left( 1+ n\pi \frac{2\pi \lambda^2 M}{{\bar{k}}^3} \right)^{-1}.
\end{equation}
Notice that, despite being apparently of order $\lambda^2$, the correction to unity is given by the ratio between powers of two small quantities, namely the effective coupling constant $\lambda/M$, and the wavenumbers ratio $\bar{k}/M$. The resulting number can be of order one,
even at small coupling constants. Observe that the probability vanishes like $\bar{k}^3/n$ at very small $\bar{k}$: this behavior is physically motivated by the fact that, as the energy approaches the cutoff, the distance between the atoms must increase to infinity in all bound states. Let us finally analyze the energy density of the electromagnetic fields. Neglecting the exponentially suppressed contribution of the square-root cuts, the energy density turns out to be related to the Fourier transform of the photon amplitude,
\begin{eqnarray}
\widetilde{\varphi}_n(x) & = & \int \frac{d k}{2\pi} \varphi_n(k) \e^{\i kx} \nonumber \\ 
& \simeq & \frac{\lambda c_A^{(n)} 2M}{\sqrt{2\pi E}} \int d k \frac{1-(-1)^n \e^{-\i kd_n}}{{\bar{k}}^2 - k^2} \e^{\i kx},
\end{eqnarray}
as 
\begin{equation}\label{energydensity}
\mathcal{E}_n(x)  \simeq  E |\widetilde{\varphi}_n(x)|^2 
 \simeq  \Bigl( \frac{2\sqrt{\pi}\lambda M}{\bar{k}} \Bigr)^2 \!p_n \sin(\bar{k}x)^2, 
\end{equation}
for $x\in [0,d_n]$, and $\mathcal{E}_n(x)\simeq 0$ outside.
Thus, the field is confined between the two atoms, and modulated with periodicity $\pi/\bar{k}$, with nodes at the positions of the atoms which act as mirrors. 
This explains the occurrence of such bound states for discrete values~(\ref{kbar}) of the interatomic distance.

\begin{figure}
\centering
\includegraphics[width=0.45\textwidth]{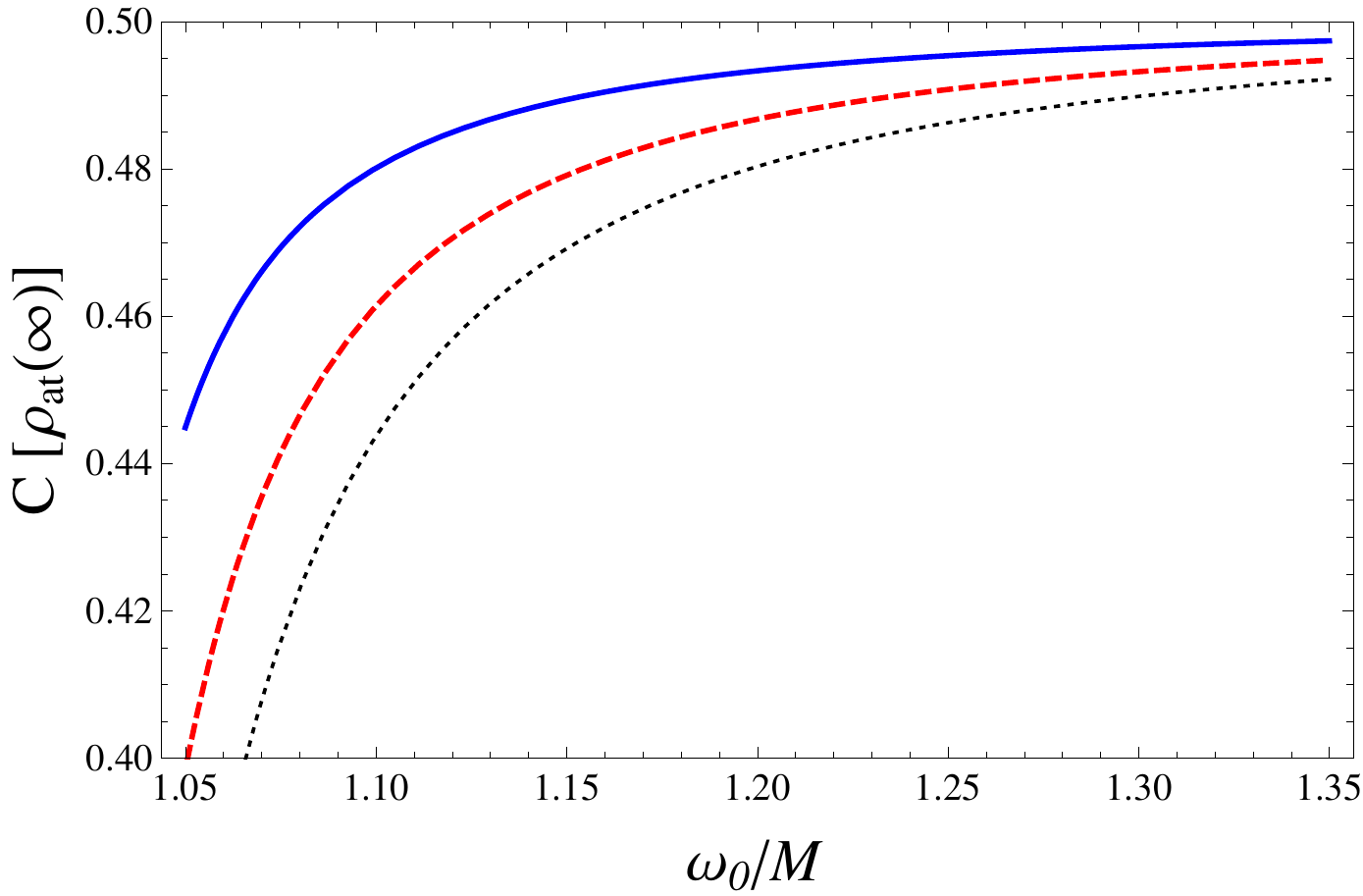}
\caption{Behavior of the concurrence $C=p_n^2/2$ of the asymptotic states $\rho_{\mathrm{at}}(\infty)$ as a function of the atomic excitation energy, for $\lambda=10^{-2}M$ and a factorized initial state. The solid (blue) line, dashed (red) line and dotted (black) line are referred to the resonant states with $n=1,2,3$, respectively.
}\label{fig:concurrence}
\end{figure}

Moreover, the structure of the bound state is
\begin{equation}\label{boundstate}
\ket{\psi_n} = \sqrt{p_n} \ket{\Psi^s}\otimes \ket{\mathrm{vac}} 
+ \ket{g_A,g_B}\otimes \ket{\varphi_n},
\end{equation}
where $s=(-1)^{n+1}$ and
$\ket{\Psi^{\pm}}=(\ket{e_A,g_B}\pm\ket{g_A,e_B})/\sqrt{2}$ are (maximally entangled) Bell states. 
This is a key feature which enables entanglement generation by atom-photon interaction. Indeed, suppose that $d=d_n$: a factorized initial state, say $\ket{\psi(0)}=\ket{e_A,g_B}\otimes\ket{\mathrm{vac}}$, can be expanded into a ``stable'' and a ``decaying'' part as
\begin{equation}\label{eAgB}
\ket{e_A,g_B;\mathrm{vac}} = \sqrt{\frac{p_n}{2}} \ket{\psi_n} + 
\sqrt{1-\frac{p_n}{2}} \ket{\psi_n^{\perp}},
\end{equation}
with $\bra{\psi_n^{\perp}}\psi_n\rangle=0$. After a transient of the order of $\ket{\psi_n^{\perp}}$'s lifetime (see discussion in the following), 
the atomic density matrix $\rho_{\mathrm{at}}(t):=\mathrm{Tr}_{\mathrm{field}} \ket{\psi(t)} \bra{\psi(t)}$ approaches
\begin{equation}
\rho_{\mathrm{at}}(\infty) =  \frac{p_n^2}{2} \ket{\Psi^s}\bra{\Psi^s} 
+ \Bigl( 1- \frac{p_n^2}{2} \Bigr) \ket{g_A,g_B}\bra{g_A,g_B},
\end{equation}
in which the atoms  have a finite probability, determined by~(\ref{probability}), to be maximally entangled.
In Figure~\ref{fig:concurrence} we display the atomic entanglement in the asymptotic state, as measured by the concurrence \cite{Wootters}. However, one could also measure the photon state and obtain, with a finite probability, a maximally entangled atomic state. The strategy is therefore the following: one prepares a factorized state, and measures whether a photon is emitted. If (after a few lifetimes) no photon has been observed, the atomic state is projected over the maximally entangled Bell state $|\Psi^s\rangle$. This can be achieved with higher probabilities for larger values of $\omega_0$. In realistic scenarios this simplified picture is challenged by the presence of losses, such that it is no longer possible to prepare an exact Bell state. Nevertheless, if losses occur on sufficiently long timescales (as compared to the decay rate of the fast pole --- see section~\ref{resolvent} below), and provided the detector efficiency is high enough, it remains possible to achieve high fidelity with a Bell state.

\section{Time evolution and bound state stability} \label{resolvent}
Let us now study the general evolution of an initial state in the atomic  sector $\mathcal{N}_{\mathrm{at}}=1$. We will use the resolvent formalism \cite{CT,NNP} to illustrate that the system relaxes towards the bound state, and to quantify the robustness of the bound state against small variations in the model parameters (such as the $A$-$B$ distance). We remark that the usefulness of the resolvent formalism goes beyond the analysis of stable states, in that it provides crucial information on the relevant timescales of the problem. Indeed, the entanglement-by-relaxation protocol described in the previous section relies on the fast decay of the unstable Bell state. The analysis of the resolvent enables to determine the lifetime of this unstable state, which must be much shorter than the typical timescales of waveguide or atomic losses, as well as the inevitably finite lifetime of the bound state (due for example to imperfect control of the A-B distance). Whenever these conditions are met, the effectiveness of the protocol is guaranteed and a long-lived entangled state may be prepared by relaxation.

The resolvent ${\cal G}(z)= (z-H)^{-1}$, with $z$ the complex energy variable, has singularities only on the real axis (on the first Riemann sheet) and the study of additional singularities (on the other Riemann sheets) yields crucial information about the dynamical stability  of the system: in particular, a pole with a non-vanishing imaginary component signals a decay process. The resolvent approach yields results that are consistent with those obtained from the analysis of the Laplace transform of the time evolution \cite{refereeA2}.

For $\lambda=0$, the free resolvent ${\cal G}_0(z)= (z-H_0)^{-1}$ has a pole on the real axis, at $z=\omega_0$, corresponding to the excited states of atoms $A$ or $B$. When interaction is turned on, this singularity splits into two simple poles, which generally migrate into the second Riemann sheet. We shall see from a non-perturbative analysis that, under resonance conditions, one of the poles falls on the real axis (and is therefore very long-lived), while the other one has a very short lifetime. Let $G(z)$ and $G_0(z)$ be the restrictions to the $\mathcal{N}_{\mathrm{at}}=1$ sector of the interacting and free resolvent, respectively.  In the basis $\{\ket{e_A,g_B},\ket{g_A,e_B}\}$ one gets
\begin{equation}
G_0(z) = \frac{1}{z-\omega_0} \left(\begin{matrix} 1 & 0 \\ 0 & 1 \end{matrix} \right) 
\end{equation}
and 
\begin{equation}
G(z)= [G_0(z)^{-1}-\lambda^2 \Sigma(z)]^{-1}= [z-\omega_0 -\lambda^2 \Sigma(z)]^{-1},
\label{eq:selfenergy}
\end{equation} 
where 
\begin{equation}
\Sigma(z) = \left(\begin{matrix} \Sigma_{AA}(z) & \Sigma_{AB}(z) \\ \Sigma_{BA}(z) & \Sigma_{BB}(z) \end{matrix} \right)
\end{equation}
is called self energy.

The resolvent $G(z)$ is analytic in the whole complex energy plane, except at  points on the real axis that belong to the spectrum of the Hamiltonian $H$. In particular it exhibits simple poles at the eigenvalues of $H$ and cuts along its continuous spectrum~\cite{CT,NNP}.
However, it can happen
that some complex poles show up on the second Riemann sheet, through the analytic continuation of the resolvent  $G_{\mathrm{II}}(z)$ from the upper half-plane to the lower half-plane under the cut~\cite{NNP,thesis}. These poles physically correspond to unstable states with energy and decay rates given by their real and imaginary part, respectively.

The particular form of the interaction Hamiltonian $V$ in~(\ref{hamiltonian}) enables one to exactly evaluate
the  self energy: 
\begin{eqnarray}
\Sigma_{AA}(z) = \Sigma_{BB}(z) & = & 
\int d k \frac{1}{\omega(k)(z-\omega(k))}, \\
\Sigma_{AB}(z) = \Sigma_{BA}(z) & = & 
\int d k \frac{\cos(kd)}{\omega(k)(z-\omega(k))}.
\end{eqnarray}
Due to the bare energy degeneracy and the symmetric structure of the self energy, the propagator can be diagonalized as
\begin{equation}
G(z) = \frac{\ket{\Psi^+}\bra{\Psi^+}}{z-\omega_0-\lambda^2 \Sigma_{+}(z)} + \frac{\ket{\Psi^-}\bra{\Psi^-}}{z-\omega_0-\lambda^2 \Sigma_{-}(z)},
\end{equation}
where
\begin{equation}\label{integralSE}
\Sigma_{s} (z) = 2 \int_M^{\infty} d\omega  \frac{\kappa_{s}(\omega)}{z-\omega}, \qquad s=\pm1,
\end{equation} 
with spectral densities
\begin{equation}
\kappa_{\pm}(E) = 
\frac{1 \pm \cos(\sqrt{E^2-M^2} d)}{\sqrt{E^2-M^2}} \chi_{[M,\infty)(E)}.
\label{eq:spectraldensity}
\end{equation}
The self-energy functions $\Sigma_{\pm}(z)$ are analytic in the cut complex energy plane  $\mathbb{C}\setminus [M,+\infty)$ and have a purely imaginary discontinuity across the cut proportional to the spectral density:
\begin{equation}
 \Sigma_{s}(E-\i 0^+) - \Sigma_{s}(E+\i 0^+) = 2 \pi \i \kappa_{s}(E).
\label{eq:discontinuity}
\end{equation}

During the continuation process into the
second Riemann sheet through the cut, the
self energy~(\ref{integralSE}) will thus get an
additional term
\begin{equation}
\Sigma_s(z) \longrightarrow \Sigma_s^{\mathrm{II}}(z) =
\Sigma_s(z)-2\pi i\kappa_s(z), \qquad z\in\mathbb{C}.
\label{eq:SigintII}
\end{equation}
Note that the new term has in general a nonvanishing imaginary
part and  is the analytical continuation of the 
discontinuity of the self-energy function across the cut.
Now, a pole 
\begin{equation}
z_{\rm p} =E_{\mathrm{p}} - \i \gamma_{\mathrm{p}}/2
\label{eq:poledef}
\end{equation} 
of $G(z)$ on the second sheet must satisfy the equation
\begin{equation}
z_{\mathrm{p}}=\omega_0+\lambda^2\Sigma_s^{\mathrm{II}}(z_{\mathrm{p}}),
\label{eq:1equpol}
\end{equation}
for $s=+1$ or $s=-1$,
where $\Sigma_s^{\mathrm{II}}(z)$ is the branch
(\ref{eq:SigintII}) of the self energy  in the second
sheet. By plugging~(\ref{eq:SigintII}) and (\ref{eq:spectraldensity}) into~(\ref{eq:1equpol}) we get
\begin{equation}\label{poles}
z_{\mathrm{p}} = \omega_0 + \lambda^2 \left( \Sigma_{\pm} (z_{\mathrm{p}}) - 4\pi \i \frac{1\pm \cos\left(\sqrt{z_{\mathrm{p}}^2-M^2} d\right)}{\sqrt{z_{\mathrm{p}}^2-M^2} } \right).
\end{equation}
It is evident from~(\ref{poles}) how the energetic degeneracy at $\lambda=0$ is lifted by interactions. Notice the presence of an imaginary component, detecting decay.

The last ingredient we need in order to get a closed expression for the complex energy poles is the evaluation of $\Sigma_s(z)$ in~(\ref{poles}). Thus, let us rewrite~(\ref{integralSE}) as an integral over $k$:
\begin{equation}\label{integralSEk}
\Sigma_{\pm} (z) = \int_{-\infty}^{+\infty} \frac{dk}{\sqrt{k^2+M^2}} \frac{1\pm e^{\i kd}}{z - \sqrt{k^2+M^2}} .
\end{equation} 
The integrand function can be analytically continued to the complex $k$ plane using the principal determination of the square root, which has nonnegative real part for all values of its argument, and is characterized by a branch cut for $k^2 + M^2 <0$ , that is
\begin{equation}
k= \pm \i \chi, \quad \text{with } \chi \in (M,\infty).
\end{equation}
Two first-order poles, symmetric with respect to the origin of the $k$ plane, are also present whenever $\Re(z)>0$:
\begin{equation}
k = \pm k_0(z) = \pm \sqrt{z^2-M^2}.
\end{equation}

\begin{figure}
\includegraphics[width=0.45\textwidth]{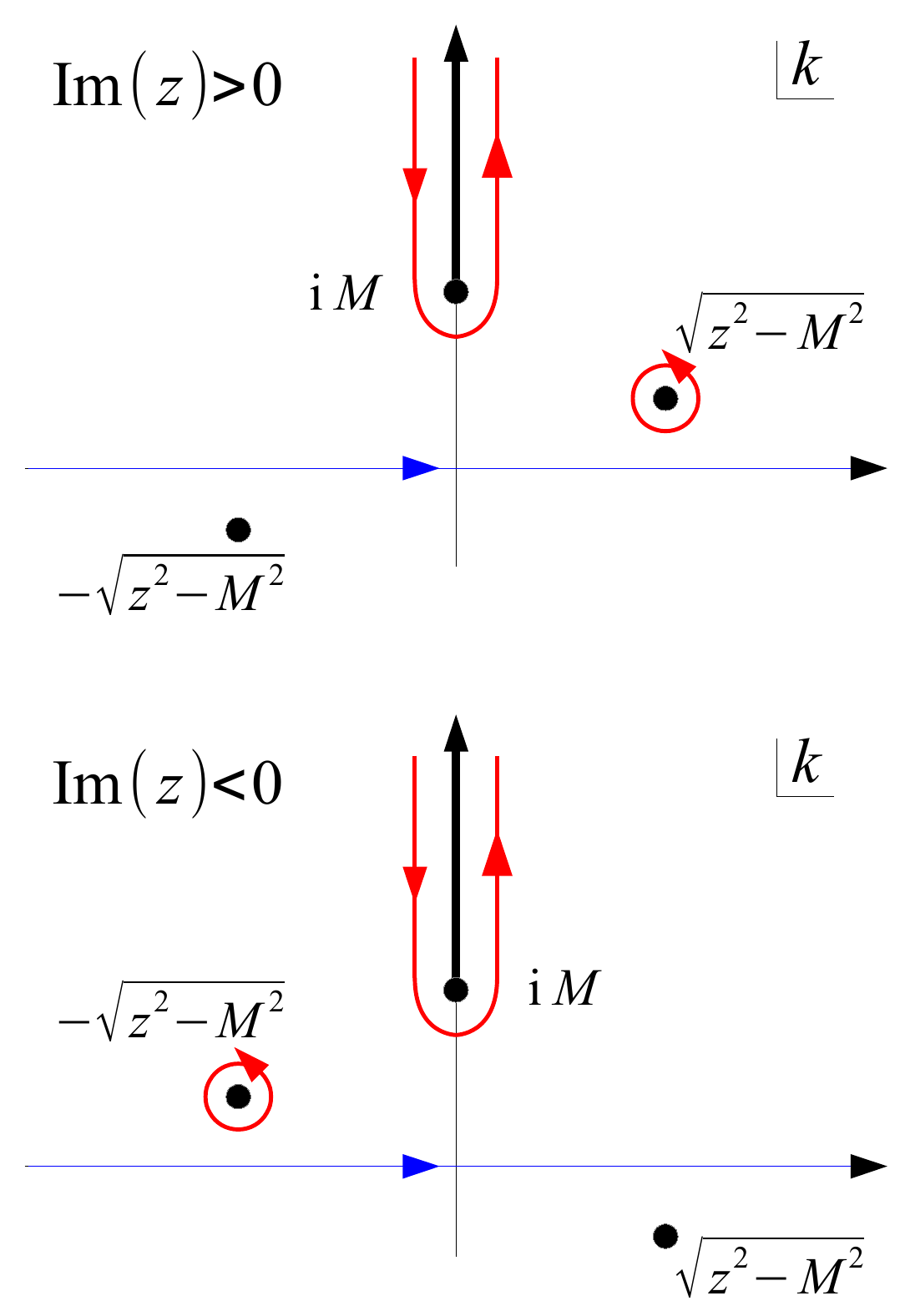}
\caption{Integration contours (red) in the complex $k$ plane for the computation of the integral in Eq.~(\ref{integralSE}) with $\mathrm{Im}(z)>0$ (upper panel) and $\mathrm{Im}(z)<0$ (lower panel).}\label{fig:integrals}
\end{figure}

By deforming the integration contours as in Figure \ref{fig:integrals} and applying Jordan's theorem, $\Sigma_{s}$ is split in two terms 
\begin{equation}
\Sigma_{s}(z) = \Sigma_{s}^{\mathrm{cut}}(z) +  \Sigma_{s}^{\mathrm{pole}}(z).
\label{eq:split}
\end{equation}
coming from the upper branch cut and from one of the two poles (see Figure \ref{fig:integrals}). Specifically, when $\mathrm{Im}(z)>0$, the pole $k_0(z)$ lies in the upper half plane, and the integral involves the residue
\begin{eqnarray}
\Sigma_{\pm}^{\mathrm{pole}}(z) &=& 2\pi \i \lim_{k\to k_0(z)} \frac{(k-k_0(z))(1\pm e^{\i kd})}{\sqrt{k^2+M^2} (z - \sqrt{k^2+M^2})} \nonumber \\
&=& - 2\pi \i \frac{1\pm e^{\i\sqrt{z^2-M^2} d}}{ \sqrt{z^2-M^2} }.
\label{eq:26}
\end{eqnarray}
Instead, when $\mathrm{Im}(z)<0$, the deformed contour in the upper plane encircles $-k_0(z)$, where the residue yields 
\begin{eqnarray}
\Sigma_{\pm}^{\mathrm{pole}}(z)  &=& 2\pi \i \lim_{k\to -k_0(z)} \frac{(k+k_0(z))(1\pm e^{\i kd})}{\sqrt{k^2+M^2} (z - \sqrt{k^2+M^2})} \nonumber \\
&=&2\pi \i \frac{1\pm e^{-\i\sqrt{z^2-M^2} d}}{ \sqrt{z^2-M^2} }.
\label{eq:27}
\end{eqnarray}

Finally,
the integrals along the cut read
\begin{eqnarray}
\label{eq:Sigmacut}
\Sigma_{\pm}^{\mathrm{cut}} (z) &=& 2 z \int_{M}^{\infty} d\chi \frac{1\pm e^{-\chi d}}{ \sqrt{\chi^2-M^2} (z^2+\chi^2-M^2) } \nonumber \\
&=& \frac{2 }{\sqrt{z^2-M^2}} \mathrm{Log} \left( \frac{z + \sqrt{z^2-M^2}}{M} \right) \pm O(e^{-Md}), \nonumber \\
\end{eqnarray}
where the contribution from $e^{-\chi d}$, that is not amenable to an explicit closed form in terms of simple functions, is nevertheless suppressed like $e^{-Md}$ and can be neglected for large values of $Md$.

\begin{figure}
\centering
\includegraphics[width=0.45\textwidth]{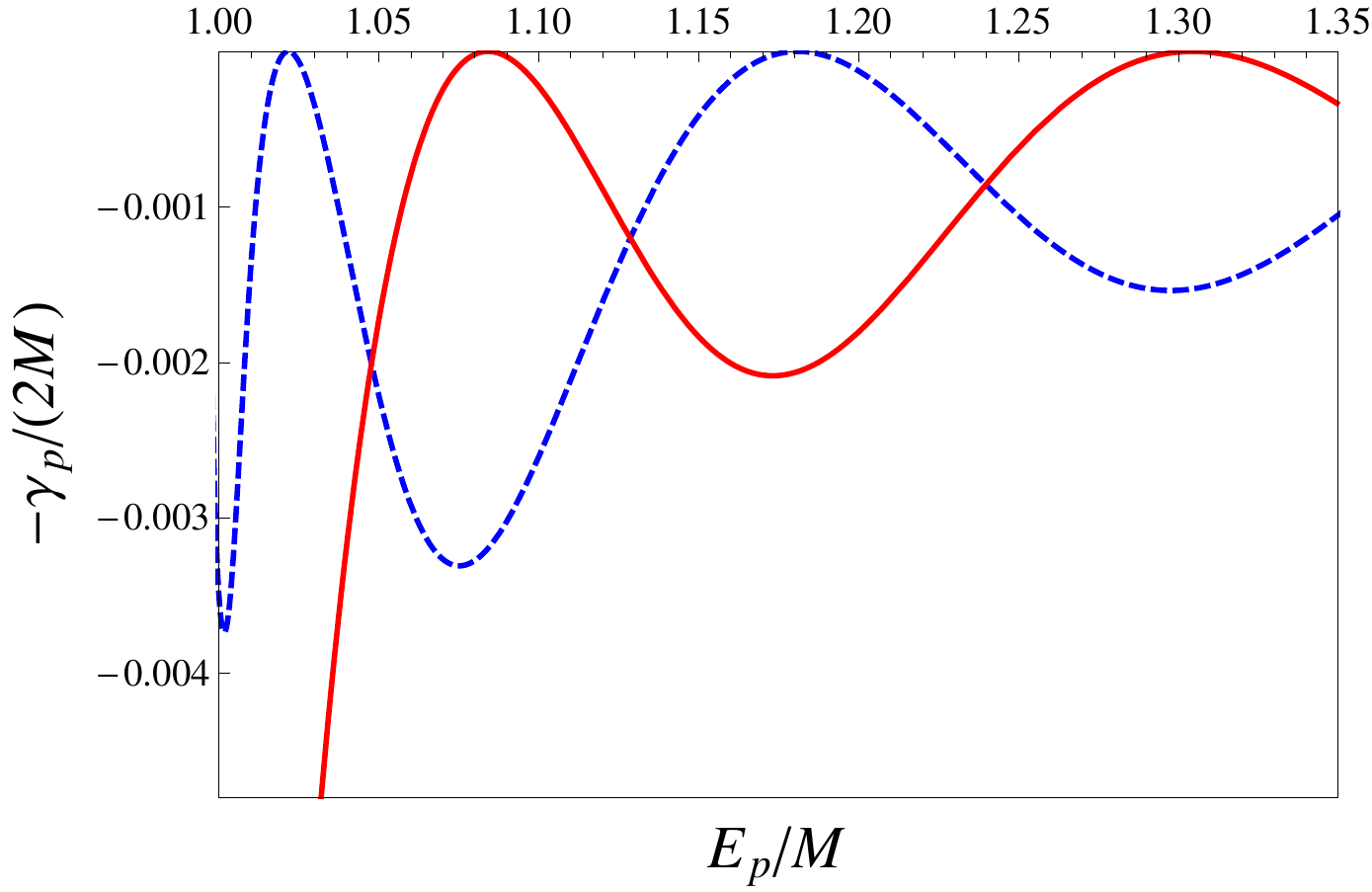}
\caption{Trajectories of the poles $E_{\mathrm{p}}^{(+)}- \i \gamma_{\mathrm{p}}^{(+)}/2$ (dashed blue line) and $E_{\mathrm{p}}^{(-)}- \i \gamma_{\mathrm{p}}^{(-)}/2$ (solid red line) on the second Riemann sheet of the complex energy plane, for $Md=15$ and $\lambda=10^{-2}M$, with varying $.95 \leq \omega_0/M \leq 1.35$. The trajectories are tangent to the real axis (they touch it whenever condition~(\ref{kbar}) is satisfied), showing that the approximate bound states are robust against variation of $\omega_0$.
Notice that the behaviour of both poles becomes non perturbative in $\lambda$ as $\omega_0 \sim M$.
}\label{fig:poles}
\end{figure}

We are now able to recognize the real resonant poles discussed in the first part of the paper as special solutions of Eq.~(\ref{poles}). Indeed, assuming that the complex energy pole~(\ref{eq:poledef})
is far from the branching point $z=M$ and that its imaginary part is almost vanishing, one can decouple the real and imaginary parts of~(\ref{poles}) and obtain from Eqs.~(\ref{eq:split})-(\ref{eq:Sigmacut})
\begin{align}
E_{\mathrm{p}} \simeq & \, \omega_0 + \frac{2 \lambda^2}{k_{\mathrm{p}}} \mathrm{Log} \left( \frac{E_{\mathrm{p}} + k_{\mathrm{p}}}{M} \right) 
\pm 2 \pi \lambda^2 \frac{\sin(k_{\mathrm{p}} d)}{k_{\mathrm{p}}}, 
\label{real} \\
\gamma_{\mathrm{p}} \simeq & \, 4\pi\lambda^2 \frac{1\pm\cos(k_{\mathrm{p}}d)}{k_{\mathrm{p}}},\label{im}
\end{align}
where $k_{\mathrm{p}} =\sqrt{E_{\mathrm{p}}^2-M^2}$.
Hence, we find that the poles in the second Riemann sheet have a cyclic behavior with respect to $d$. This result is in agreement with the one obtained in Ref.~\cite{refereeA1} in the Markovian approximation. In particular, when $d=d_n$, as defined in Eq.~(\ref{kbar}), the real part of the pole equations is solved by $E_{\mathrm{p}}=\sqrt{\bar{k}^2+M^2}$. In this case, one of the poles corresponds to the entangled bound state, and has vanishing imaginary part, while the other signals an unstable state with associated decay rate 
\begin{equation}\label{gammau}
	\gamma_{\mathrm{p}}^{(u)}=8\pi\lambda^2/\bar{k}
\end{equation}
Even if, strictly speaking, bound states only occur for discrete values of $d$, it can be readily checked that while the energy shift is linear, the decay rate of the stable pole is quadratic for $d\to d_n$
\begin{equation}\label{gammap}
\gamma_{\mathrm{p}}^{(s)} 
\simeq 2 \pi \lambda^2 \bar{k} (d-d_n)^2, 
\end{equation}
implying that the state $\ket{\psi_n}$ remains very long-lived close to resonance. Eq.~\eqref{gammap} quantifies the robustness of the bound states against variations of the parameter $d$. Note how Eqs.~\eqref{gammau} and \eqref{gammap} provide essential information on the feasibility and effectiveness of the entanglement generation protocol: firstly, it is necessary that the condition $\gamma_{\mathrm{p}}^{(s)}\ll\gamma_{\mathrm{p}}^{(u)}$ is satisfied, which is equivalent to the condition $\bar k^2(d-d_n)^2\ll1$. Secondly, $\gamma_{\mathrm{p}}^{(u)}$ must be much larger than any decay rate associated with loss processes (e.g. waveguide losses). Even though approximate analytical expressions such as Eqs.~\eqref{gammau} and \eqref{gammap} are extremely valuable, we emphasize that our methodology is capable of capturing the {\it exact} behaviour of the poles against variations in the model parameters. To illustrate this, in Fig.~\ref{fig:poles} we show the trajectories of the poles~(\ref{eq:poledef}) in the complex energy plane, obtained by fixing $M$ and $d$ and varying the bare excitation energy $\omega_0$. On the one hand, we are thus able to assess quantitatively the robustness of bound states against variations in $\omega_0$. On the other hand, Fig.~\ref{fig:poles} demonstrates how our methodology allows one to interpolate seamlessly between perturbative and non-perturbative regimes.

\section{Off-resonant bound states}\label{offres}

Let us briefly discuss the behavior of bound states with $E<M$. In this case, the atoms are not expected to decay. Nonetheless, they interact by coupling to the evanescent modes of the waveguide. Scrutiny of Eqs.~(\ref{eigenvalueA})-(\ref{eigenvalueB}) shows that there are bound states for all $d$, whose energy satisfies
\begin{equation}\label{Esmall}
\left(
\begin{array}{cc}
E-\omega_0-\alpha(E) & - \beta(E) \\ -\beta(E) & E-\omega_0-\alpha(E)
\end{array}\right)
\left( \begin{array}{c} c_A \\ c_B \end{array} \right)
=0,
\end{equation}
with
\begin{align}
\alpha(E) & = - \frac{\lambda^2}{\sqrt{M^2-E^2}} \!\left(\pi + 2 \arctan\! \frac{E}{\sqrt{M^2-E^2}} \right)\! , \\
\beta(E) & = - \frac{\lambda^2}{\sqrt{M^2-E^2}} \, 2\pi e^{-{\sqrt{M^2-E^2}}d} .
\end{align}
where we have again neglected the $O(e^{-Md})$ contributions from branch-cut integration in the complex $k$ plane. If the coupling is small and the excitation energy $\omega_0$ is far from the threshold $M$ for photon emission, the above equations reduce to an effective Hamiltonian eigenvalue equation in the $\mathcal{N}_{\mathrm{at}}=1$ sector. The eigenvalues read
\begin{equation}
E^{(\pm)} = \omega_0 + \alpha(\omega_0) \pm \beta(\omega_0),
\end{equation}
with $c_A=c_B$ for the plus sign (ground state) and $c_A=-c_B$ for the minus sign. These bound states are not associated to any resonance. It is also possible to check that the electromagnetic energy density falls like $\exp(-\sqrt{M^2-E^2}|x|)$ away from the atoms. 

Since the eigenstates of the effective Hamiltonian are Bell states, the evolution of an initially factorized state is characterized by oscillations between two orthogonal maximally entangled states with period $2\pi/\beta(\omega)$. Compared to entanglement by relaxation, this mechanism yields unit concurrence \cite{refereeA2}. On the other hand, the process can be very slow, since the energy splitting is exponentially suppressed with the interatomic distance, and requires the fine tuning of an optimal time to stop the interactions, which is not required in the spontaneous entanglement process described in Section \ref{model}.

The case $E\to M$ is more interesting, since the physics becomes nonperturbative. Equations (\ref{eigenvalueA})-(\ref{eigenvalueB}) admit a singlet and a triplet solution. For the singlet case, the bound state with $E=M$ is obtained at a finite excitation energy: 
\begin{equation}
\omega_0 = M - \frac{2\lambda^2}{M} + 2\pi\lambda^2 d,
\end{equation}
in which, due to a cancellation of divergences, the correction to the bare energy is still perturbative in $\lambda^2$. This singlet solution approximates the dark eigenstate $\ket{\Psi^-}\otimes\ket{\mathrm{vac}}$
occurring at $d=0$. The triplet, instead, survives as a real eigenstate even for $\omega_0\geq M$. However, since $c_A=c_B$ implies that the integrals over the field become divergent in this limit, the population in the $\mathcal{N}_{\mathrm{at}}=1$ sector is suppressed to fulfill normalization, and the contribution of this pole to the expansion~(\ref{eAgB}) can be safely neglected.

\section{Extension to generic dispersion relations}\label{sec:general}
While we have worked out in detail the case of a rectangular waveguide, we emphasize that our methods can be applied to a generic
dispersion relation $\omega(k)$.
We start by noticing that Eqs.~\eqref{eigenvalueA}-\eqref{eigenvalueF} lead in full generality to the implicit condition
\begin{equation}\label{generalE}
	E = \omega_0 + \lambda^2 \int d k \frac{1-(-1)^n \e^{-\i kd}}{\omega(k)(E-\omega(k))},
\end{equation}
which must be satisfied by the bound state energy $E$. For the existence of a resonant (i.e. above threshold) bound state, it is evident that also the condition $\bar{k}d=n\pi,\, n\in\mathbb{N}$ must hold, where $\bar k>0$ satisfies $E=\omega(\pm\bar{k})$. If this were not the case, the right hand side of Eq.~\eqref{generalE} would diverge.
We assume that such $\bar k$ exists and is unique. This is the case, for example, when $\omega(k)$ is an increasing lower-bounded function of $|k|$.  
Moreover, the possibility of non-resonant eigenstates below threshold follows as in the case of a rectangular waveguide.

Moving on to the complex energy plane, the analysis of poles proceeds along the same lines as in section~\ref{resolvent}, albeit the existence of compact analytical expressions will rely on the specific functional form of $\omega(k)$. The pole contribution to the self energies can be generalized by replacing
the denominators in Eqs.~(\ref{eq:26})-(\ref{eq:27}) with $\omega(k_0)\omega'(k_0)$, and the square root in the exponentials with $k_0$. In the perturbative regime, this change does not affect formally the ratio of the decay rates of stable and unstable poles close to a resonance, namely [see Eqs.~\eqref{gammau}-\eqref{gammap}]
\begin{equation}\label{Gratio}
\frac{\gamma_{\mathrm{p}}^{(s)}}{\gamma_{\mathrm{p}}^{(u)}} = \frac{1}{4} \bar k^2(d-d_n)^2,
\end{equation}
where $d_n=n\pi/\bar k$. We can see that quantitative differences between models arise in the inversion of the dispersion relation as a function of the energy. The quantity in Eq.~\eqref{Gratio} gives a clear indication of the potential of a given model to generate entanglement by relaxation. While losses would inevitably degrade the quality of the achievable entangled state, Eq.~\eqref{Gratio} may be seen as posing a fundamental limit to the entangling capabilities of a given system, a limit which would persist even in an idealized lossless scenario.

\section{Conclusions and outlook} \label{conclusions}
We analyzed stable and unstable states of a pair of atoms in a waveguide, finding that an entangled bound state exists for discrete values of the interatomic distance. This implies that an initially factorized atomic state can spontaneously relax towards a long-lived entangled state. By analyzing the poles of the resolvent operator, we have shown how to quantify the robustness of the entangled bound state to small variations in the model parameters, and how to identify the timescales that are crucial for the preparation of an entangled state by relaxation.\\ 
While it has been pointed out that quantum computation may be achievable in waveguide-QED trough effective photon-photon interactions \cite{barangio}, focusing on the atomic degrees of freedom may also hold significant potential for applications in Quantum Information \cite{decofree}. Further investigation will thus be devoted to the analysis of many-atom systems \cite{leo1,leo2,leo3,leo4}, in which photon-mediated interactions could possibly produce stable configurations such as W states or cluster states.

\section*{Acknowledgments}
We thank F. Ciccarello for useful discussions. MSK thanks the UK EPSRC, the Royal Society and the FP7 Marie Curie programme (grant number 317232).
PF was partially supported by the Italian National Group of Mathematical Physics (GNFM-INdAM).
PF, SP and FVP are partially supported by INFN through the project ``QUANTUM".

\appendix
\section{Derivation of the quasi-1D free field Hamiltonian}
\label{1DH}

We derive here the Hamiltonian in Eq.~(1) of the main text from first principles. Let us consider a waveguide of infinite length, parallel to the $x$ axis, characterized by a rectangular cross section with $y\in [0,L_y]$ and $z\in [0,L_z]$. We conventionally assume that $L_y>L_z$. A common choice is $L_y/L_z=2$. In a generic guide made of a linear dielectric with uniform density and coated by a conducting material, the boundary conditions for the electric and magnetic fields on the surface $S$ read
\begin{equation}
E_x |_S = 0 \qquad \text{and} \qquad \left. \frac{\partial B_x}{\partial n} \right|_S = 0,
\end{equation} 
with $\partial/\partial n$ denoting the normal derivative with respect to the surface. Transverse electric (TE) modes are characterized by $E_x=0$ everywhere in the guide and obtained by imposing $\partial B_x /\partial n = 0$ on the surface. On the other hand, transverse magnetic (TM) modes have $B_x=0$ identically. If the waveguide is rectangular, the boundary conditions for TE modes reduce to
\begin{equation}
\left. \frac{\partial B_x}{\partial y} \right|_{y=0} = \left. \frac{\partial B_x}{\partial y} \right|_{y=L_y} = \left. \frac{\partial B_x}{\partial z} \right|_{z=0} = \left. \frac{\partial B_x}{\partial z} \right|_{z=L_z} = 0,
\end{equation}
which limits the form of the longitudinal magnetic field to the real part of
\begin{equation}
B_x = B_0 \cos\left( \frac{m \pi y}{L_y} \right) \cos\left( \frac{n \pi z}{L_z} \right) e^{i (k x - \omega_{m,n}(k) t)},
\end{equation}
with $m,n\in \mathbb{N}^2 \backslash \{(0,0)\}$ and $B_0$ a constant. \\
The integers $m$ and $n$ label the mode $\mathrm{TE}_{m,n}$. The dispersion relation with respect to the longitudinal momentum has the same form as a massive relativistic particle, 
\begin{equation}
\omega_{m,n}(k) = \sqrt{(v k)^2 + \omega_{m,n}(0)^2},
\end{equation}
with $\omega_{m,n}(0) = v \left[ \left( \frac{m \pi y}{L_y} \right)^2 + \left( \frac{n \pi z}{L_z} \right)^2 \right]^{\frac{1}{2}}$,
where the mass term is called the \textit{cutoff frequency} of the mode, and $v=(\mu\epsilon)^{-1/2}$ is the phase velocity in the waveguide, assumed isotropic and nondispersive with magnetic permeability $\mu$ and dielectric constant $\epsilon$. Since $L_y<L_z$, the $\mathrm{TE}_{1,0}$ mode has the lowest cutoff frequency. It can be proved \cite{jackson} that $\omega_{1,0}(0)$ is also lower than the cutoffs of all TM modes. Thus, at sufficiently low energy the contribution of the higher energy modes can be neglected, and propagation occurs effectively in one dimension.

The $\mathrm{TE}_{1,0}$ mode is characterized by the following behavior of the fields
\begin{eqnarray}
B_x &=& B_0 \cos\left( \frac{\pi y}{L_y} \right) e^{i (k x - \omega_{1,0}(k) t)}, \\
B_y &=& - i\frac{k L_y B_0}{\pi} \sin\left( \frac{\pi y}{L_y} \right) e^{i (k x - \omega_{1,0}(k) t)}, \\
E_z &=& i\frac{\omega_{1,0}(k) L_y B_0}{\pi} \sin\left( \frac{\pi y}{L_y} \right) e^{i (k x - \omega_{1,0}(k) t)},
\end{eqnarray}
with the other three components vanishing. These fields can be derived from the (transverse) vector potential
\begin{equation}
A_z = \frac{L_y B_0}{\pi} \sin\left( \frac{\pi y}{L_y} \right) e^{i (k x - \omega_{1,0}(k) t)}.
\end{equation}
The mode can be quantized by introducing the time-0 field operators
\begin{eqnarray}
{\bm{A}}^{(1,0)}(\bm{r}) & = & \int dk \left( \frac{\hbar}{2\pi \epsilon \omega_{1,0}(k) L_y L_z} \right)^{\frac{1}{2}} \sin\left( \frac{\pi y}{L_y} \right) \nonumber \\
& & \qquad \times \left[ a(k) e^{ikx} + a^{\dagger}(k) e^{-ikx} \right] \hat{u}_z, \\
{\bm{E}}^{(1,0)}(\bm{r}) & = & i \int dk \left( \frac{\hbar \omega_{1,0}(k)}{2\pi \epsilon L_y L_z} \right)^{\frac{1}{2}} \sin\left( \frac{\pi y}{L_y} \right) \nonumber \\
& & \qquad \times \left[ a(k) e^{ikx} - a^{\dagger}(k) e^{-ikx} \right] \hat{u}_z,
\end{eqnarray}
with $[a(k),a^{\dagger}(k')]=\delta(k-k')$ and $\hat{u}_z=(0,0,1)$. The electric field energy operator associated to the mode thus reads
\begin{eqnarray}
\mathcal{E}_{el}^{(1,0)} &=& \frac{\epsilon}{2} \int d\bm{r} :\!\left(E_z^{(1,0)}(\bm{r})\right)^2\!: \nonumber  \\
&=& \frac{1}{2} \int dk \,\hbar \omega_{1,0}(k) \Bigl[a^{\dagger}(k)a(k)  \nonumber \\
& & \qquad - \frac{a(k) a(-k) + a^{\dagger}(k) a^{\dagger}(-k)}{2} \Bigr]
\end{eqnarray}
with $:(...):$ denoting normal ordering, while the magnetic field energy can be evaluated using the relation $\bm{B}^{(1,0)} = \bm{\nabla}\times\bm{A}^{(1,0)}$:
\begin{eqnarray}
\mathcal{E}_{mag}^{(1,0)} &=& \frac{\epsilon}{2} \int d\bm{r} :\!\left(\partial_y A_z^{(1,0)}(\bm{r})\right)^2\! + \!\left(-\partial_x A_z^{(1,0)}(\bm{r})\right)^2\!: \nonumber \\
&=& \frac{1}{2} \int dk \,\hbar \omega_{1,0}(k) \Bigl[a^{\dagger}(k)a(k) \nonumber \\
& & \qquad+ \frac{a(k) a(-k) + a^{\dagger}(k) a^{\dagger}(-k)}{2} \Bigr].
\end{eqnarray}
Thus, the free Hamiltonian for the electromagnetic field takes the diagonal form
\begin{eqnarray}
H^{(1,0)} &=& \mathcal{E}_{el}^{(1,0)} + \mathcal{E}_{mag}^{(1,0)} \nonumber \\
&=& \int dk\, \hbar \omega_{1,0}(k) a^{\dagger}(k)a(k) \nonumber \\
&=& \hbar v \int dk\, \sqrt{k^2+\left(\frac{\pi}{L_y}\right)^2} a^{\dagger}(k)a(k).
\end{eqnarray}
It is worth noticing that the analogy with a massive boson is not limited to the dispersion relation. Indeed, the quantum theory of the mode can be mapped onto a real scalar theory in one dimension, by introducing the operators
\begin{eqnarray}
\alpha(x) &=& \int dx\, \sqrt{\frac{\hbar}{2(2\pi)\omega_{1,0}(k)}} \left[ a(k) e^{ikx} + a^{\dagger}(k) e^{-ikx} \right], \nonumber \\
\Pi(x) &=&-i \int dx\, \sqrt{\frac{\hbar\omega_{1,0}(k)}{2(2\pi)}} \left[ a(k) e^{ikx} - a^{\dagger}(k) e^{-ikx} \right], \nonumber \\
\end{eqnarray} 
satisfying
\begin{eqnarray}
[\alpha(x), \Pi(x')] = i \hbar \delta(x-x')
\end{eqnarray}
and related to the vector potential and the electric field by multiplication. The Hamiltonian can be expressed in terms of the field operator $\alpha(x)$ and its canonically conjugated momentum $\Pi(x')$ as
\begin{eqnarray}
H^{(1,0)} &=& \frac{1}{2} \int dx : \!\Bigl[ \left( \Pi(x) \right)^2 + v^2 \left( \partial_x \alpha(x) \right)^2 \nonumber \\
& & \qquad+ v^4 \!\left(\frac{M}{\hbar}\Bigr)^2 \left( \partial_x \alpha(x) \right)^2 \right] \!: 
\end{eqnarray}
with $ M:= \frac{\pi\hbar}{v L_y}$, which also allows to identify a linear Hamiltonian density $\mathcal{H}(x)$ such that $H^{(1,0)} = \int dx \mathcal{H}$.

\section{Interaction Hamiltonian}

The interaction between the field and an artificial atom, made up of a particle trapped in a potential $V(\bm{r})$, can be obtained by  the minimal coupling prescription:
\begin{eqnarray}
H_{\mathrm{at}} &=& \frac{1}{2m_e} \left( \bm{p} - e \bm{A}^{(1,0)}(\bm{r}) \right)^2 + V(\bm{r}) \nonumber \\
 &=& H_{\mathrm{at}}^0 - \frac{e}{m_e} \bm{p}\cdot\bm{A}^{(1,0)}(\bm{r}) + \frac{e^2}{2m_e} \left( \bm{A}^{(1,0)}(\bm{r}) \right)^2 , \nonumber \\
\end{eqnarray} 
with $\bm{r}$ and $\bm{p}$ the canonically conjugated position and momentum of the artificial ``electron''. The transverse choice $\bm{\nabla}\cdot\bm{A}= 0$ for the vector potential makes the ordering with respect to $\bm{p}$ immaterial. We adopt a two-level approximation for the atom, retaining only the ground state $\ket{g}$ and the first excited state $\ket{e}$, satisfying
\begin{equation}
H_{\mathrm{at}}^0 \ket{g} = 0, \qquad H_{\mathrm{at}}^0 \ket{e} = \hbar \omega_0 \ket{e}.
\end{equation}
Furthermore, we apply long-wavelength approximations to the interaction terms, which enable one to neglect the $O(e^2)$ contribution, whose relevance is suppressed like the ratio of the photon momentum to the particle momentum \cite{CT}, and to apply a dipolar approximation to the $O(e)$ term. The position operator $\bm{r}$ is replaced by a non dynamical center-of-mass position $\bm{r}_0$. The interaction Hamiltonian thus reads
\begin{eqnarray}
H_{\mathrm{int}}^{(dip)} &=& - \frac{e}{m_e} A_z^{(1,0)}(\bm{r}_0) \Bigl[ \bra{g}p_z\ket{g} \ket{g}\bra{g} + \bra{e}p_z\ket{e} \ket{e}\bra{e} \nonumber \\
& &  + \bra{g}p_z\ket{e} \ket{g}\bra{e} + \bra{e}p_z\ket{g} \ket{e}\bra{g} \Bigr] . 
\end{eqnarray}
The assumption that the expectation value of momentum vanishes in the eigenstates of the free Hamiltonian simplifies the interaction. Moreover, the canonical commutation relation can be used to obtain
\begin{eqnarray}
\bra{e}p_z\ket{g} &=& \frac{i m}{\hbar} \bra{e} [H_{\mathrm{at}}^0,z] \ket{g} = i m \omega_0 \bra{e}z\ket{g} =: i m \omega z_{eg} \nonumber \\
&=& i m \omega_0 |z_{eg}| e^{i\theta_{eg}},
\end{eqnarray}
by which the mass $m_e$ disappears from the theory, and the Hamiltonian takes the form of a coupling between the dipole moment $D_{eg}=e|z_{eg}|$ and the electric field. Finally, we can define new canonically conjugated field operators $b(k):= e^{-i(\theta_{eg}+\pi/2)} a(k)$ and retain only the rotating-wave terms $b(k)\ket{e}\bra{g}$ and $b^{\dagger}(k)\ket{g}\bra{e}$, to obtain the interaction operator
\begin{eqnarray}
H_{\mathrm{int}}^{(dip,RW)} &=& \omega_0 D_{eg} \left( \frac{\hbar}{2\pi\epsilon v L_y L_z} \right)^{\frac{1}{2}} \int \frac{dk}{(k^2+(vM/\hbar)^2)^{1/4}} \nonumber \\
& & \qquad \times \left[b(k) \ket{e}\bra{g} e^{ikx_0} + b^{\dagger}(k) \ket{g}\bra{e} e^{-ikx_0} \right]. \nonumber \\
\end{eqnarray}
Notice that $y_0=L_y/2$ has been used. The dynamics for the atom pair is thus determined by
\begin{equation}
H = H_{\mathrm{at},A}^0 + H_{\mathrm{at},B}^0 + H^{(1,0)} + H_{\mathrm{int},A}^{(dip,RW)} +  H_{\mathrm{int},B}^{(dip,RW)}
\end{equation}
with atom $A$ in $x_0=0$ and atom $B$ in $x_0=d$.

\section{Energy density}

The study in the main text has been focused on the $\mathcal{N}=1$ sector, spanned by the wavefunctions
\begin{eqnarray}
\ket{\psi_1} &=& c_A \ket{e_A,g_B; vac} + c_B \ket{g_A,e_B; vac} \nonumber \\
&& + \int dk F(k) \ket{g_A,g_B;k}.
\end{eqnarray}
Using the scalar Hamiltonian density defined in Section \ref{1DH}, one can compute the energy density
\begin{eqnarray}
\bra{\psi_1} \mathcal{H}(x) \ket{\psi_1} & = & \frac{1}{2} \Bigl[ \bra{\psi_1}:\left( \Pi(x) \right)^2:\ket{\psi_1} \nonumber \\
&& \qquad + v^2 \bra{\psi_1}:\left( \partial_x \alpha(x) \right)^2:\ket{\psi_1} \nonumber \\
&& \qquad + v^4 \!\left(\frac{M}{\hbar}\right)^2 \bra{\psi_1}:\left( \partial_x \alpha(x) \right)^2:\ket{\psi_1} \Bigr] \nonumber \\
& = & \left| \int dk \sqrt{\frac{\hbar\omega_{1,0}(k)}{2(2\pi)}} F(k) e^{ikx} \right|^2 \nonumber \\
&& + \left| \int dk \frac{\hbar v k}{\sqrt{2(2\pi)\hbar\omega_{1,0}(k)}} F(k) e^{ikx} \right|^2 \nonumber \\
&& + \left| \int dk \frac{v^2 M}{\sqrt{2(2\pi)\hbar \omega_{1,0}(k)}} F(k) e^{ikx} \right|^2 . \nonumber \\
\end{eqnarray}
This stucture can be simplified if one assumes that the dominant contribution to the integrals comes from the poles of $F(k)\sim A_{+} (k-k_0)^{-1} + A_{-} (k+k_0)^{-1}$. Neglecting the corrections yielded by square-root branch-cut integration, one obtains
\begin{eqnarray}
\bra{\psi_1} \mathcal{H}(x) \ket{\psi_1} & \simeq & \hbar \omega_{1,0}(k_0) \left| \int \frac{dk}{2\pi} F(k) e^{ikx} \right|^2 \nonumber \\
&=:& \hbar \omega_{1,0}(k_0) \left| \tilde{F}(x) \right|^2,
\end{eqnarray}
which is used to compute the energy density for the resonant states.

\end{document}